\documentclass[12pt]{article}		
\usepackage{fancyhdr}
\usepackage{multicol}
\usepackage{latexsym}
\usepackage{graphicx}
\usepackage[T1]{fontenc}			
\usepackage{setspace}				
\usepackage{marvosym}				
\usepackage{seqsplit}				
\usepackage[hyphens]{url}
\urldef\myurlgamechanger\url{https://www.dla.mil/Portals/104/Documents/DLMS/Summit/Resources/GAMECHANGER_Slick%20Sheet_2024.pdf}

\usepackage[round,semicolon,authoryear]{natbib}	
\bibliographystyle{chicago}

\usepackage{authblk} 

\begin{document}
	

\title{Stop Saying ``AI''}
\author[1,2,3]{Nathan Wood \thanks{Authors' contributions: NGW conceived of the article, served as initial author for Sections 1 and 6, integrated the work of all the authors, and served as article lead. SR conducted primary research and writing of Section 2. EZB conducted primary research and writing of Section 4. AvW conducted primary research for Section 3. MB conducted primary research for Section 5. HB oversaw the discussion of AI systems and challenges, providing technical expertise and guidance. DKS aided project development overall. All authors participated in revision of the article and ensuring overall cohesion. }}
\author[4]{Scott Robbins}
\author[1]{Eduardo Zegarra Berodt}
\author[1]{Anton Graf von Westerholt}
\author[1,5]{Michelle Behrndt}
\author[1]{Hauke Budig}
\author[1]{Daniel Kloock-Schreiber}
\affil[1]{Institute of Air Transportation Systems, Hamburg University of Technology}
\affil[2]{Ethics + Emerging Sciences Group, California Polytechnic State University San Luis Obispo}
\affil[3]{Center for Environmental and Technology Ethics -- Prague}
\affil[4]{Academy for Responsible Research, Teaching, and Innovation, Karlsruhe Institute of Technology}
\affil[5]{Department of Philosophy, University of Hamburg}
\date{\today}

\maketitle	


\begin{abstract}

Across academia, industry, and government, ``AI'' has become central in research and development, regulatory debates, and promises of ever faster and more capable decision-making and action. In numerous domains, especially safety-critical ones, there are significant concerns over how ``AI'' may affect decision-making, responsibility, or the likelihood of mistakes (to name only a few categories of critique). However, for most critiques, the target is generally ``AI'', a broad term admitting many (types of) systems used for a variety of tasks and each coming with its own set of limitations, challenges, and potential use cases. In this article, we focus on the military domain as a case study and present both a loose enumerative taxonomy of systems captured under the umbrella term ``military AI'', as well as discussion of the challenges of each. In doing so, we highlight that critiques of one (type of) system will not always transfer to other (types of) systems. Building on this, we argue that in order for debates to move forward fruitfully, it is imperative that the discussions be made more precise and that ``AI'' be excised from debates to the extent possible. Researchers, developers, and policy-makers should make clear exactly what systems they have in mind and what possible benefits and risks attend the deployment of those particular systems. While we focus on AI in the military as an exemplar for the overall trends in discussions of ``AI'', the argument's conclusions are broad and have import for discussions of AI across a host of domains. 

\bigskip
	

\end{abstract}


\section{Introduction}\label{sec_intro}

``Artificial intelligence'' is currently a household topic, with academia, industry, governments, and civil society at large all weighing in. Some tout all that may be achieved by using ``AI'', but there are many worries, critiques, and objections to its widespread use in certain domains or for certain (types of) tasks. Regardless of one's own level of optimism or pessimism about AI, a common element in the debates is for discussion to center around ``artificial intelligence'' or ``AI'' in some general sense, and few discussions pay heed to the extraordinary breadth of systems that are or contain ``AI'' in them and are used in various domains.\footnote{Notable exceptions to this are \cite{meerveld2023irresponsibility,meerveld2025operationalising,boulanin2024risks,barretttaylor2025ai,goncharuk2025chasing,knack2025defence,voloshchuk2025artificial}.} More than this, ``AI'' is often used as a shorthand to indicate either one very specific (type of) system under question\footnote{E.g., \cite{chubb2022speeding,guest2025against,suzuki2025people}.} or is used in a diffuse and imprecise way,\footnote{E.g., \cite{schuller2019artificial,schwarz2020humanity,black2024strategic,davidovic2025rethinking}.} often with it being left unclear which situation one is in. 

In this article, we focus on the military domain and provide a loose enumerative taxonomy of various different types of AI systems that can or may soon be found in militaries, examining what those systems might be used for and what challenges attend their use. This loose taxonomy is not exhaustive, but rather provides a broad sketch of various forms of ``military AI'', along with an accounting of the potential issues they may raise. The main purpose of this article is to make clear that at no point should we be discussing, critiquing, or attempting to regulate ``AI'' as such, be it AI in the military, medical, or any other field, as there falls under this umbrella such a wide breadth of systems, even when the discussion has been restricted to a specific domain. Rather, academic, policy, and regulatory debates ought to focus on discrete (types of) AI systems and the applications they are designed and used for. We close by highlighting what has been succinctly stated by Paul Scharre, that ``AI is not a discrete technology, like railroads or airplanes. It is a general-purpose enabling technology, like electricity, computers, or the internal combustion engine, with many applications''.\footnote{\cite{scharre2023four}, p. 3.} This view of AI not as \textit{a discrete technology} but rather \textit{an enabling technology} is critical to bear in mind, as the question is then not whether AI, in all its potential instantiations, is acceptable in the military, medical, or any other domain, but rather whether discrete particular instantiations are. 

The article is structured as follows. We begin by examining AI systems used in decision support roles (Section \ref{sec_decisionsupport}), looking to what sorts of AI systems or capabilities contribute to this task and how they may be used. We also highlight distinct challenges attending AI-enabled decision support systems (AI-DSS) used for intelligence processing (Section \ref{subsec_decisionsupport_ISR}) and for target nominations or recommendations (Section \ref{subsec_decisionsupport_targetnomination}). We then move to an examination of AI in weapon systems (Section \ref{sec_autonomousengagement}), focusing on autonomous weapon systems (AWS) and the potential role of AI in so-called ``last mile targeting'' for systems which have only limited autonomy or which lose connection to human operators (Section \ref{subsec_autonomousengagement_lastmile}), for controlling systems which are wholly autonomous throughout an engagement (Section \ref{subsec_autonomousengagement_fullautonomy}), and for controlling swarms of systems or munitions acting in concert (Section \ref{subsec_autonomousengagement_swarms}). Pulling back from autonomous systems, we then examine how AI may be used in support of operations and engagements involving both manned and unmanned systems (Section \ref{sec_optimization}), exploring AI systems' potential use and attendant challenges in logistics roles (Section \ref{subsec_optimization_logistics}), in support of missions (Section \ref{subsec_optimization_support}), and in the development of new tactics and doctrine (Section \ref{subsec_optimization_r&d}). Continuing further back from the line of contact, we explore how AI may be used in support functions or in streamlining the bureaucratic elements of military operations (Section \ref{sec_bureaucraticAI}), looking to how systems may be designed to improve institutional or doctrinal functions and understandings (Section \ref{subsec_bureaucraticAI_HR}) and how it may affect mundane aspects of the military as a corporate body (Section \ref{subsec_bureaucraticAI_corporatestreamlining}). Finally, we summarize our findings and argue for the need for specificity going forward in the debates on ``military AI'' and ``AI'' more generally (Section \ref{sec_specificity}). 

However, before moving onto the core discussion, it is necessary to clarify certain key terms and limits. First, with the article's focus on ``AI'', it will be useful to have a working definition of the concept to avoid misunderstanding. This is especially relevant given the recent generalized hype around ``AI'' and the use of the term for virtually any and every system using algorithmic processes or automation of complex tasks.\footnote{\cite{tucker2022artifice} provides useful and succinct discussion of some of these trends, along with concrete policy responses that may be taken to combat these. For further discussion of AI hype across numerous domains, see the 2024 special issue of \textit{AI and Ethics} on ``The Ethical Implications of AI Hype: Examining the overinflation and misrepresentation of AI capabilities and performance''.} 

In defining ``artificial intelligence'', one may set a high bar, demanding ``the same scope of intelligence as we see in human action''.\footnote{\cite{newell1976computer}, p. 116.} Alternatively, one may present a more general view, holding that an ``AI'' system must possess ``methods of achieving goals in situations in which the information available has a certain complex character''\footnote{\cite{mccarthy1988mathematical}, p. 308.} or must have a capacity ``to adapt to its environment while operating with insufficient knowledge and resources''.\footnote{\cite{wang2019defining}, p. 19. See also \citep{wang1995non}, the 2020 special issue of the \emph{Journal of Artificial General Intelligence} dedicated to discussing Wang's view (Volume 11, Issue 2), and the EU AI Act, esp. p. 39.} More colloquially, we may follow Timnit Gebru, founder and executive director of the Distributed AI Research Institute and former co-lead of the Ethical AI research team at Google, who stated that her ``understanding of the field is that you try to create machines or things that can do more than what's been programmed into them''.\footnote{\cite{marx2023fall}.} Taking all these points into consideration, in what follows, we use Luciano Floridi's definition which itself follows the European Union and United States views, that 

\begin{quote}
	Artificial Intelligence (AI) refers to an engineered system that can, for a given set of human-defined objectives, generate outputs –- such as content, predictions, recommendations, or decisions –- learn from data, improve its own behaviour, and influence people and environments.\footnote{\cite{floridi2023brussels}, p. 9. Note that Floridi's definition ``is not a scientific definition but a legal one'' (\textit{ibid}), but given that most critical debates around AI are focused on ethical and legal concerns and not scientific ones, Floridi's conception is fitting. An alternative but influential conception (especially in computer science and robotics communities) is Stuart Russell and Peter Norvig's view which, according to them, has become ``so pervasive that we might call it the \textbf{standard model}'', namely that AI is ``\textit{the study and construction of agents that \textbf{do the right thing}}'', where ``[w]hat counts as the right thing is defined by the objective that we provide to the agent'' (\cite{russell2021artificial}, p. 22, emphasis in original). Despite its widespread use in technical disciplines, Russell and Norvig's use of ``right thing'' is apt to confuse discussions of ethical or legal aspects of AI development and use, and is moreover arguably over-inclusive, in that a toaster ``does the right thing'' when it properly toasts bread, though that system is clearly nothing like an artificial intelligence. For these reasons, among others, we defer to Floridi's conception.} 
\end{quote}

The second point worth clarifying is that, for reasons of article length, our focus here is on \emph{systems} in particular, with less emphasis on specific deployments of those systems or the significant impact that context can have on the ethical and legal use of systems. This is worth making explicit, as rarely will a system itself indicate anything about its potential to be permissibly used, with context and parameters or limitations of use usually holding far more sway. For purposes of this taxonomical work, we must unfortunately set aside such more practical ``system $+$ context'' examinations. The work here though establishes core baselines for pursuing such further work, and complements recent research looking more closely at how contexts impact on deployment decisions.\footnote{See especially \cite{meerveld2025operationalising}. See also \cite{sipri2020limits,kwik2024controlling,gcreaim2025responsible,trabucco2025control2,trabucco2025control3}.}

To forestall certain objections, we also stipulate already that this article will not develop a complete or proper taxonomy, in terms of developing general principles of classification for AI-enabled systems in the military (or any other domain).\footnote{\cite{webster2025taxonomy}.} Rather, due to the sheer number of systems which may incorporate AI and the fact that many systems will incorporate a variety of AI-enabled processes or (sub-)systems, often with overlap, we focus on a loose enumerative categorization of relevantly similar systems. This approach is in keeping with our aim to provide a pragmatic examination of systems and their challenges, rather than to develop a rigorous classificatory schema for AI-enabled systems and their envisioned missions.\footnote{For examples of such classificatory work, see, e.g., \cite{samoili2020ai,gill2025edge}.}

It is also important to stress that for any cutting-edge technology, and especially those in the military domain, many aspects of system design and development may be classified or only partially publicized. As such, discussion of some AI techniques, capabilities, and systems will require educated guesswork or deduction from existing knowledge about needs and capabilities. Throughout this article, we have prioritized examples of systems where more information is available, but inevitably, some systems which are shrouded in more secrecy required discussion. For those, we have extrapolated from existing information the most likely uses of AI, drawing out the potential challenges associated with these. 

Finally, we would highlight that while we focus here on AI-enabled systems in the military domain specifically, our general findings bear relevance for virtually all discussions of AI. Moreover, that there is such a breadth of systems and potential deployments in just this one domain -- the military -- indicates just how widely AI may be used, with attendantly broad (sets of) challenges. Thus, as we will show, specificity is crucial for debates and discussions to be fruitful, and this is not limited to departments of defense. Across every domain where AI is being explored, touted, or marketed, we must be more specific about what we are talking about, as the challenges are not germane to ``AI'' as such, but rather to specific systems, their design, their limitations, and what we (can) do about those limitations. With that in mind, let us examine some of the very many ways ``AI'' is or soon may be used in the military, remembering that the military represents only one of many domains where AI may be deployed. 

\section{Decision Support}\label{sec_decisionsupport} 

A central use of AI in the military is for supporting intelligence, planning, and decision-making. Decision support systems (DSS) -- model-based platforms that process data to assist decision-makers -- have been in use since the 1970s, with early DSS relying heavily on intricate decision trees which were gradually enhanced by embedding additional domain knowledge within their structures.\footnote{\cite{liao2000case,davis2005implications}.} The value and impact of DSS may be illustrated by examining the Observe-Orient-Decide-Act loop (OODA loop), a decision-making framework introduced by Colonel Jon Boyd.\footnote{\cite{enck2012ooda}.} Within the OODA loop, DSS contribute primarily by parsing and analyzing raw intelligence, building coherent operational pictures from disparate data to support the observe and orient phases. Subsequently, DSS can generate, evaluate, and rank alternative courses of action to streamline the decide phase, offering decision-makers a structured set of options aligning with the assessed context.\footnote{DSS are also deployed in numerous other domains, most notably in medicine. See, e.g., \cite{sim2001clinical,castaneda2015clinical}.\label{ftnote_dss_medicine}}

Recent advances in artificial intelligence -- particularly machine-learning techniques -- have shifted DSS development from manual crafting of rule sets toward data-driven training. By exposing systems to large volumes of relevant data, developers allow DSS to ``learn'' patterns that can be used to generate insights and recommendations, thereby improving the quality and relevance of the support provided to decision-makers. AI's broad applicability also means that it may be incorporated into DSS in various ways: deep-learning models for visual perception can be used to recognize objects such as weapons, soldiers, tanks, aircraft, and other critical assets in real-time sensor feeds; deep-learning models can enhance signal analysis across a range of data types such as imagery, video, audio, thermal, seismic, radar, and other communications -- including social-media posts, telephone calls, radio broadcasts, and emails -- to compile actionable intelligence; and large language models (LLMs) are advertised to generate clear and concise briefings, draft reports, or produce narrative explanations, giving analysts automated text-generation tools that adapt to evolving operational pictures. AI-enabled DSS (AI-DSS) thus promise to greatly expand the capabilities of traditional decision support systems, and do so in a way that can potentially allow for re-training as needed in fluid environments.\footnote{For explication of core aspects of AI-DSS, as well as both critical and optimistic views of AI-DSS, see, e.g., \cite{dear2019artificial,klonowska2020article,ivanov2023automated,boulanin2024risks,nadibaidze2024ai,sipri2025comparison,reynolds2025speed}. In line with note \ref{ftnote_dss_medicine} above, AI-DSS are also becoming relevant for domains outside the military, especially in medicine. See \cite{antoniadi2021current,shiang2022artificial,asiri2024ai}.}

However (AI-)DSS also raise a number of critical challenges. Fundamentally, the impact of framing effects means (AI-)DSS may undermine judgment and careful consideration of variables when making decisions,\footnote{\cite{tversky1981framing}.} and basic cognitive limitations in humans can make (AI-)DSS potentially dangerous to deploy in certain settings.\footnote{\cite{phillips2020decision}.} A more common objection is that (AI-)DSS may exacerbate the effects of numerous cognitive biases humans have (notably, automation bias and action bias), leading to misuse or poor use of such systems.\footnote{\cite{walrath1989aiding,arnott2006cognitive,cummings2017automation,phillips2019cognitive}.} Within the defense domain specifically, there are concerns that, among other things, (AI-)DSS may cause more civilians to be harmed than would otherwise be the case,\footnote{\cite{stewart2023algorithms}.} they may undermine humans' ability to ``meaningfully control'' military systems,\footnote{See, e.g., \cite{binns2018reducing,schwarz2021autonomous,robbins2023many,robbins2025what}. See also \cite{wood2026extensions} for a related critique of (AI-)DSS.} and they may even create command challenges and potentially undermine strategic thinking.\footnote{See \cite{hunter2023never,johnson2023automating,barretttaylor2025ai} for exploration of (AI-)DSS and command challenges, especially at the strategic level. \cite{johnson2020delegating} examines uses of AI-DSS which may be destabilizing and therefore potentially highly dangerous. See also \cite{lushenko2026ai} for recent empirical analysis of the impact of AI on strategic-level decision-making in the United States military.} Some of these problems have seen proposed solutions,\footnote{E.g., \cite{allik2024framework,dorsch2024explainable} provide both design-oriented and institutional approaches. See also \cite{fritz2024deference} for a general defense of the view that we should defer to (AI-)DSS under certain conditions.} but there remain many open questions regarding the responsible, ethical, and legal use of AI-DSS in the military domain. To highlight these challenges, we will briefly present a few concrete examples of AI-DSS currently in development or in use for both intelligence processing and advisement purposes and for target nominations and recommendations. 

\subsection{Intelligence Processing}\label{subsec_decisionsupport_ISR}

Modern intelligence, surveillance, and reconnaissance (ISR) systems can capture immense amounts of data, and for the wealth of information coming from satellites, drones, sensors, radar, etc. to be useful, it must be processed and turned into actionable guidance. Traditionally, human analysts did this work, and with the advent of modern computing, some information could be pre-parsed or processed with the aid of automated systems. AI-enabled systems are now aiding with this as well, allowing for more sophisticated and incisive processing to be done at scale, a point which is becoming increasingly unavoidable given the scale of modern information gathering. Some AI-enabled systems are able to integrate data from numerous sources and fuse them into concrete intelligence products, while other systems may be tailored to highly specific use cases, with each raising distinct challenges.

Palantir's MetaConstellation is a data-integration platform that brings together information from a range of sensors, e.g., satellites, radars, acoustic detectors, maritime automatic identification systems, or even social media posts. Using this data, machine-learning (ML) models then extract usable signals: computer-vision models analyze visual inputs to detect and track objects (vehicles, structures, terrain changes); natural-language processing systems parse textual feeds for usable intelligence on hostile actors, locations, intent, etc.; and other ML-based systems operate on the various data streams being fed into MetaConstellation. Palantir claims that using the processed data, MetaConstellation can also provide automatic notifications when certain user-defined rule-based triggers are satisfied, allowing for human operators to tailor the system to distinct operational and tactical needs. Palantir adds this is further underpinned by a built-in natural-language query interface within MetaConstellation which allows users to pose questions in plain English (or any other supported language) in order to extract further information on recommendations, elicit further guidance, or gain clarification on system outputs.\footnote{See \url{https://www.palantir.com/offerings/metaconstellation/}. Note also that though Palantir (and its various products) are regular features in news stories, journalists often seem to echo the marketing of Palantir without further scrutiny. See, e.g., \cite{macaes2023palantir}.} As such, it brings together a variety of AI capabilities, from ML-based systems used to analyze, classify, and generate advisement from various data sources to natural-language processing (NLP) which is used for both user interface and textual queries as well as for analysis of open-source intelligence and intercepted communications. 

AI-DSS can come in narrower forms as well. One such system is provided by Clearview AI, which claims to have compiled 2 billion images of faces collected from Russian social media (and 10 billion images overall), and using these, has developed an AI-based facial recognition system.\footnote{\url{https://clearview.ai}.} In developing their system, Clearview AI relied on scraping images from the internet, a practice of uncertain legality,\footnote{\cite{hill2024clearview}.} and the system's overall reliability also remains unclear.\footnote{\cite{clayton2023clearview}.} Despite these points, Clearview AI has offered its services to Ukraine, and though the exact use cases of this system have not been specified, it would likely be used for identifying the dead or potential enemy combatants, aiding with checkpoint screenings, and debunking misinformation on social media.

A similarly narrow AI system used to aid intelligence processing and decision-making is the audio analysis system developed and marketed by Primer.\footnote{\url{https://primer.ai/primer-command/}.} According to company marketing, Primer uses ``analytical machine learning'' (likely natural language processing) to analyze audio signals intelligence, with the system being able to scan for, for example, Russian language, place names, people, weapons, etc., and then summarize the information into succinct bullet points.\footnote{\cite{gourley2022new}.} The system is also claimed to be able to filter out music, non-Russian audio, and other interference, taking the isolated speech and automatically translating it into English and providing a summary of the translation.\footnote{\textit{Ibid}.}

All of the above systems use ``AI'' in some fashion, but each system makes use of AI in (sometimes radically) different ways, drawing out distinct potential challenges. Issues of bias and false positives, which are common in the general discourse around AI, are highly relevant and potentially prevalent for a system like Clearview AI's facial recognition platform. However, Primer's audio analysis system brings forward very few concerns around potential ethnic biases related to how one looks, instead raising possible issues relating to system reliability or how well the system has been trained on local speech mannerisms or expressions; some local dialects may admit expressions which are anodyne in the region, but could be flagged as suspect by those unfamiliar with the local culture, and a poorly trained system thus may make mistakes with such manners of speech. Similar concerns hold for virtually all languages. A much more extensive system like Palantir's MetaConstellation brings forward a host of other distinct issues, ranging from how reliable the system is at base to how the user interface may tap into and exacerbate cognitive biases such as automation and action bias. MetaConstellation's incorporation of a variety of ML-based systems also makes it liable to numerous distinct challenges rooted in these varying applications of AI. MetaConstellation's ability to provide recommendations and notifications on potential tactical or operational threats also makes it liable to be a core contributor to kinetic strikes, a fact which makes any challenges or limitations to the system all the more critical. 

\subsection{Target Nomination/Recommendation}\label{subsec_decisionsupport_targetnomination}

Considering how brittle AI systems can be,\footnote{\cite{heaven2019why}.} their use for recommendation or nomination of potential targets in kinetic strikes is apt to raise particularly significant ethical and legal challenges. Even so, there are numerous systems currently in use or under development precisely for this task. 

One of the most notable and widely reported AI-DSS used for target nominations or recommendations is the Lavender system utilized by the Israel Defense Forces (IDF). Essentially a database cataloging human subjects, entries in Lavender are created and prioritized by analyzing behavioral ``features'' drawn from open-source and signals-intelligence sources: patterns of phone calls and messaging metadata, social media connections, and frequent changes of residence are combined to generate a risk score for each person. The system then produces targeting recommendations that rank individuals according to the inferred likelihood that they pose a threat, with the highest-scored entries flagged for possible lethal action.\footnote{Information on Lavender is primarily available from reporters interviewing Israeli intelligence analysts. See, e.g., \cite{abraham2024lavender,mckernan2024israel}.} 

Complementing Lavender is Gospel, another Israeli AI-based system focused on locating and evaluating non-human targets such as buildings, bridges, tunnels, power installations, and other critical infrastructure. Gospel achieves this by processing ``drone footage, intercepted communications, surveillance data and information drawn from monitoring the movements and behavior patterns of individuals and large groups''.\footnote{\cite{davies2023gospel}.} Israel claims that a human operator must always approve any lethal or destructive action, after which the targeting coordinates are transmitted to appropriate weapon platforms, and in defending their use of the system, Israel has downplayed it, likening it to a glorified Excel sheet.\footnote{\cite{serhan2024israel}.}

The aforementioned Palantir also offers an AI-enabled system for targeting support: Titan, a military truck equipped with hardware and software to enable AI targeting without requiring access to the cloud. Described as a ``mobile ground station'', Titan is advertised as improving strike targeting and accuracy by taking in information from ``multiple domains''.\footnote{\url{https://www.palantir.com/titan/}.} Though little is known about Titan's precise functionality -- other than that it is AI-enabled and used for targeting -- two Titan vehicles were recently delivered to the United States Army.\footnote{\cite{subin2025titan}.}

The above AI-DSS used for target recommendations or nominations bring forward some challenges already sketched, in particular, potential issues related to AI's brittleness, to bias in data and system training, and the likelihood of false positives leading to mistaken and potentially illegal strikes being carried out. A distinct further challenge of AI-DSS used for targeting support is that, where systems used for intelligence processing may allow for some offloading of intellectual and analytical work during military planning, these systems would not necessarily radically supplant human judgment for the most significant decisions, like those over strikes and the potential killing of humans.\footnote{Cf. though, \cite{talbert2026triage}.} But AI-DSS used for target recommendations or nominations may do precisely this, allowing humans to offload the morally and legally weighty task of deciding upon targets, on who might live or die.\footnote{See \cite{wood2026extensions} for further discussion of the issues of ``will-offloading'' systems.} Even when the entirety of the actual decision-making rests with humans, the impact of framing effects\footnote{\cite{tversky1981framing,nelson1997toward,druckman2001evaluating,druckman2001limits}.} and how recommendations are presented may also lead humans to be predisposed to taking an AI's assessments for granted or accepting them without due credulity. For such AI-DSS, cognitive biases and design choices about how the human-machine interface functions can be highly impactful,\footnote{\cite{arnott2006cognitive,phillips2019cognitive,jimenez2024considering,minotra2024reviewing}.} making these systems much more fraught with regards to potential consequences than mere intelligence processing AI-DSS might be.\footnote{Note though that both purely intelligence-oriented and target-nominating systems may make the exact same mistakes, with the same severity and level of inaccuracy. The difference in scale of the problem is due only to the significance of outputs (intelligence analysis vs. target nomination).}

All in all, the precise manner and use of AI-DSS will greatly impact on what challenges they face, whether those challenges can be addressed, mitigated, or avoided, and whether they undermine the potential to use systems in accordance with ethical, legal, and operational norms. Moreover, as AI-DSS may be used for such varied tasks as intelligence processing or advisement or for target nominations and recommendations, they present a broad swath of potential distinct issues which are highly system- and context-specific. 

\section{Autonomous Targeting/Engagement}\label{sec_autonomousengagement} 

In addition to AI systems used to augment, complement, or potentially offset elements of human analysis and decision-making, certain AI-enabled systems are also being developed with an eye toward full system autonomy for certain tasks or missions. So-called ``autonomous weapon systems'' (AWS), that is, systems which can select and engage targets without contemporaneous human input,\footnote{\cite{icrc2021positiona}, p. 1; \cite{usdod2023directive3000.09}, p. 21. See also \cite{icrc2014autonomous}, p. 5; \cite{wood2023awsclarification}.} have existed for decades, fulfilling numerous roles.\footnote{\cite{sipri2020limits,icrc2021positionb,heller2023concept,wood2023awsclarification,wood2023awsresponsibility}.} However, where early AWS were comparatively simple and constrained, the incorporation of AI into these systems opens possibilities for ever more sophisticated, open-ended, and lengthy deployments. In general, AI-enabled autonomous weapon systems (AI-AWS) may be less apt to raise certain of the challenges sketched above -- notably, those relating to offloading decision-making\footnote{\cite{wood2026extensions}.} -- but their ability to execute processes independent of human input and sometimes even without oversight brings other potential significant risks.\footnote{For an overview, see \cite{rashid2023artificial}.} The severity and exact nature of these risks depends however on how AI-AWS are developed and deployed.

\subsection{Limited or ``Last Mile'' Autonomy}\label{subsec_autonomousengagement_lastmile}

The most constrained form of AI-AWS are those used for limited or what some call ``last mile'' autonomous engagement.\footnote{\url{https://balticviper.com/viper-last-mile-tracking/}; \url{https://thefourthlaw.ai/}; \url{https://taf-ua.com/en/products/last-mile-en/}. Another central application of ``last mile'' autonomy in military operations is for front-line logistical support. For reasons of space, we omit discussion of this use case, but the interested reader may consult, e.g., \cite{thornton2018swarming,sander2024logistics,ergozi2025logistics}.} There are two broad classes of AI use in ``last mile'' autonomous systems, with constrained autonomy embedded to meet discrete operational and tactical needs. The first is AI-AWS where the system performs final target selection and engagement, but does so in line with highly specific parameters and deployment choices human operators have already made. These systems are thus utilizing AI and autonomy to optimize targeting, extend human operators' reach, or offset other tactical and strategic needs, but they are not covering vast unbounded areas searching for targets. The second type of AI use is for giving systems autonomous capabilities in the event that communications or connections to operators are lost. Such systems utilize AI and autonomy as a fail-safe to ensure missions may still be accomplished even if interference or electronic warfare interrupt control of the system. These systems are apt to have limits to how much autonomy may be embedded and under what conditions it may kick in, but in general, they allow for humans to create contingencies in deployments of AI-AWS, in line with mission goals and the requirements of ethics and international law. 

Numerous AWS already in use fit the mold of the first type of AI-AWS sketched above, systems which make final target selections and engagements based on parameters and deployment constraints set by designers or operators in advance. From anti-radar and anti-tank systems to long-range anti-ship missiles (LRASM) capable of selecting specific targets among groups of surface vessels, autonomous target selection and engagement is not a new capability.\footnote{E.g., respectively, the U.S.-made AGM-45 Shrike or Israeli Harpy anti-radiation missiles; the Brimstone anti-tank system; and the LRASM developed by the United States Defense Advanced Research Projects Agency (DARPA) and manufactured by Lockheed Martin.} Some firms have begun incorporating AI into such standoff weapons as well, with, for example, the South Korean firm Hanwha Aerospace producing the Cheongeom anti-tank guided missile (ATGM) which is ``equipped with an artificial intelligence (AI) algorithm, which can automatically capture fixed targets in case of emergency without operator intervention through deep learning of target images of more than 800,000 frames''.\footnote{\cite{yeonhee2022development}. See also \cite{udoshi2022cheongeom,dapa2025vol140}.}

Other companies and countries have looked to (limited) autonomy as a means of responding to electronic warfare (EW) which interferes with human control of systems. In particular, Ukrainian firms have identified last mile autonomy as a critical capability for countering Russian EW, with one manufacturer, TAF Drones, offering a ``Last Mile module'' which ``enables precise targeting even if communication is lost, the system is exposed to electronic warfare, or the target is in motion''.\footnote{\cite{pokotylo2025taf,taf2025ai}. See also \cite{hrazhdan2025fourth}, reporting on a similar offering from Ukrainian firm The Fourth Law.} Such uses of AI are not properly to allow for fully autonomous weapon systems or platforms, but rather to create a possibility for autonomy when direct control breaks down, becomes less reliable, or is expected to become difficult. Such AI-AWS are thus often not going to be properly ``autonomous'' weapon systems in many instances, but will retain that capability to address emerging tactical needs. 

The challenges and potential issues of systems with last mile autonomy will generally be limited to concerns of reliability, predictability, and likelihood of mistakes in targeting.\footnote{See, e.g., \cite{sharkey2010saying,guarini2012robotic,hrw2012losing,sparrow2015twenty,winter2020compatibility}. Cf. \cite{schmitt2012autonomous,wood2026bombs} for response to some of these concerns. See also \cite{wood2024reliability} for deeper discussion of reliability, predictability, and AWS.} Thus, while AI-DSS raise numerous and varied concerns germane to the precise nature of the AI-DSS or its context of use, many of the problems of AI-AWS will be limited to base worries about how brittle the system is or how well it can be relied upon. For example, while the South Korean Cheongeom ATGM may be touted as using an algorithm trained on over 800,000 frames, it is important to be clear about just how much training this entails; standard cinema films use 24 frames per second and high-motion computer programs often have a default of 60 frames per second. While it is likely that training of the algorithm involves still frames chosen with care, putting those frames together and playing them at cinema/computer speed still only amounts to between 556--222 minutes of footage time (9.25--3.7 hours). For a system which may be critical to national defense and which is intended to destroy vehicles and likely kill their occupants, one may reasonably question whether less than ten hours of footage is sufficient for an algorithm used for targeting to be reliable. 

\subsection{Full Mission Autonomy}\label{subsec_autonomousengagement_fullautonomy}

While most AWS fulfill narrow roles or are designed and deployed to achieve discrete limited tasks, advanced AI-AWS may allow for full mission autonomy, where a system is given a broad goal and, relying on its training and internal algorithms, allowed to develop and execute a plan of attack of its own determination. While full mission autonomy may be pursued for virtually any type of AWS and any mission profile or domain of operations, a central area of current research and development for such systems is in air-to-air combat and in the support of manned aircraft.

Currently, there are numerous aerial systems under development which are intended to have full autonomy across a range of tasks, such as Boeing's MQ-28 Ghost Bat\footnote{\cite{boeing2026ghost,gordon2025australian}.} and the Turkish Bayraktar Kizilelma.\footnote{\cite{baykar2026kizilelma,naval2025kizilelma}.} For most of these so-called ``collaborative combat aircraft'' (CCA),\footnote{\cite{dimascio2025collaborative}.} the aim is generally not for them to be sent alone on missions, but rather to be sophisticated enough to act as subordinate ``wingmen'' of human air combatants who would direct wings of such AI-AWS in an area. For any CCA, and indeed any fully autonomous platform in general, AI may be incorporated into various tasks or roles such as, e.g., navigation, fuel use optimization, target identification, selection, engagement, weapons release optimization, post-strike battle-damage assessment, and post-mission analysis, with each task raising discrete ethical, legal, and potentially tactical and operational challenges. The precise nature of a fully autonomous system's challenges will, however, depend on the precise nature of the system itself, the AI(s) it incorporates, and how any algorithmic systems have been trained. 

In addition to wholly new platforms, some companies have begun developing software solutions for embedding fully autonomous AI systems into existing airframes. For example, the German firm Helsing recently tested its Centaur system on a Saab Gripen-E, with the AI system being given full control of flight and air-to-air combat against a separate piloted Gripen. Though the Centaur-controlled system had a human on board supervising the test, the AI software is reported to have been acting fully autonomously, relying on a Reinforcement Learning algorithm and able to ``learn, in only 24 hours, decades of virtual air combat experience''.\footnote{\cite{helsing2025centaur}, translated from German by the authors.}

Fully autonomous systems raise numerous potential issues, especially in how these may be incorporated into human-machine teams. This point is especially acute for the range of collaborative combat aircraft intended to act as force multipliers for air groups which centrally feature a human acting as a wing commander for increasingly AI-enabled and autonomous combat wings. As the number of systems potentially under a commander's control and the sophistication of them increase, this may create ever increasing cognitive demands on human commanders of AI-AWS. The increased tempo of AI-enabled combat may also cut against demands for tight human-machine teaming, as the human may begin to be viewed as a temporal bottleneck to combat decision-making and engagement. The pressure to reduce human input or constrain a human's potential impact on combat tempo is apt to exacerbate existing ethical and legal challenges though, making AI-AWS of this type potentially fraught. 

\subsection{Swarms}\label{subsec_autonomousengagement_swarms}

In addition to AI being incorporated to augment individual platforms or make platforms able to act as (relatively) independent subordinates of a human commander giving broader orders --  rather than operators giving explicit inputs -- AI may also serve to enable numerous platforms to act in concert with one another. AI-enabled small- or micro-drone swarms are gaining increasing attention, especially given the small-drone-intensive and EW-saturated battlefields in Ukraine.\footnote{\cite{vgi2025swarms}.} And as any future conflict between peer adversaries seems almost certain to involve use of drones and EW, the ability of AI systems to coordinate numerous platforms makes this an area of rich further study. Building on ongoing development,\footnote{E.g., \cite{quantumsystems2024swarmprogress,quantumsystems2024swarmmilestone}.} it also seems possible AI-enabled drone swarms may not be far over the horizon. 

Swarmed autonomous systems raise all of the challenges of individual AI-AWS, but may also elevate additional potential issues relating to, for example, system connectivity and how individual drones may behave when connection to the swarm or swarm leader is lost; for swarms with strike capabilities, the precise behaviors of weapon-carrying drones is apt to be of core concern. The ``reasoning'' of swarms may also be fully alien to humans, undermining possibilities for contemporaneous oversight and complicating after-action revision of deployments.\footnote{See discussion of MARL systems in Section \ref{subsec_optimization_r&d}} Due to the relative newness of these systems, it is difficult to say with certainty precisely what array of additional worries may arise, as the technology and its use cases are apt to develop in unforeseeable ways, but suffice to say, AI-enabled swarms will raise, at the least, any concerns relating to AI as such and any concerns relating to AI-AWS, in addition to swarm-specific ethical, legal, and operational challenges. 

\section{Process Optimization}\label{sec_optimization} 

Beyond advisory and direct applications for kinetic strikes, AI may also support broader operations and missions, playing a vital role and helping build the backbone of an effective military.\footnote{\cite{nato2025role}.} These applications may include the optimization of logistic chains,\footnote{\cite{lacroix2023future}.} acquisition,\footnote{\cite{usarmypa2024pilot}.} predictive maintenance,\footnote{\cite{hardin2023rapid}.} aerial refueling path planning or automation,\footnote{\cite{jann2024assisted}.} and, potentially even more significantly, for the development of new strategies and doctrine.\footnote{\cite{kim2025human,payne2025strategic}. Cf. \cite{hunter2023never,talbert2026triage}.} Yet even for non-kinetic or \textit{merely} supporting roles, integration of AI can still bring risks and challenges, as evident when one considers just a few of the non-kinetic uses of AI in support of military operations. 

\subsection{(Predictive) Maintenance}\label{subsec_optimization_logistics}

Maintenance is a critical determinant of the effectiveness, reliability, and longevity of systems. This is especially true in the military sector, where equipment failure can be the deciding factor between mission success and failure, between life and death. Maintenance is also a highly personnel-demanding enterprise,\footnote{\cite{nussbaum1978aircraft,usarmy2025fm304500}.} representing a significant running cost for modern militaries, and one which generally increases as platforms become more complex.\footnote{\cite{wolfe2022maintenance,tirpak2018retire}.} 

%
Modern systems which generate and save diverse high-fidelity data from numerous components, sub-systems, and sensors, can, however, improve maintenance schedules. With data streams covering a wide range of variables such as, e.g., propulsion thermodynamics, structural g-loads, and meteorological exposure, when synthesized with manufacturer specifications, historical maintenance logs, and field expertise, this massive volume of data can enable predictive, preventive, proactive, and reactive maintenance strategies and optimization.\footnote{\cite{ollila1999maintenance,gackowiec2019overview,moleda2023corrective,yazdi2024maintenance}.}

Following techniques already adopted in the civilian aviation sector -- such as Lufthansa Technik's Digital Twin initiative -- sensor fusion and precise log data can be used to plan preventive, predictive, and scheduled maintenance.\footnote{\cite{monroy2023lufthansa,lufthansa2023twin}.} Such efforts can also lead to maintenance which is not just responsive to needs but also proactive in preventing problems. Capitalizing on their ability to process massive amounts of data, machine learning algorithms and other AI techniques may further improve and streamline these processes by improving the accuracy of predictions and problem identification,\footnote{\cite{carvalho2019systematic}.} an effort already being undertaken by, e.g., the United States Air Force through their Predictive Analytics and Decision Assistant (PANDA).\footnote{\cite{usaf2025aidoctrine}.} Data analysis AI systems developed through machine learning thus open possibilities for improving maintenance while also potentially reducing its costs overall. 

Despite their potential benefits, there are also risks to relying on AI-enabled or assisted (predictive) maintenance which are worth bearing in mind. For one, machine learning systems may have a naturally learned survivor bias, where failures of systems which prove catastrophic may not be as well anchored in the algorithm's recommendations, because catastrophic failures may often lead to complete losses of data (and platforms) creating blind spots in the machine's training regimen.\footnote{For fuller explication of survivor bias, see, e.g., \cite{wald1980survivor,mangel1981abraham,lockwood2021fooled}.} An algorithm trained on historical data may also ``lack imagination'', making it less capable of handling novel shifts to maintenance or operational realities. In war -- or any other adversarial environment -- adversaries are apt to alter circumstances in unforeseen and sometimes unforeseeable ways, and a maintenance schedule trained only on what has happened in the past or been thought of previously will be ill-equipped to deal with shifting tactical, operational, and strategic realities without it first seeing at least some retraining efforts. The necessary scale of that retraining could also impact running operations. Finally, AI systems are apt to be incapable of considering certain strategic options and facts that may impinge on a maintenance schedule. For instance, in an attrition-centric conflict, some platforms may be determined unworthy of all but the most basic maintenance to keep them minimally combat-ready -- as they may be expected to be lost soon -- and a sophisticated and subtle algorithm might not respond well to such a radical change to its basic characteristics (potentially raising ``out of distribution'' or ``concept drift''problems). 

\subsection{Mission Support}\label{subsec_optimization_support}

In addition to maintenance and logistics, AI may also be incorporated into more direct aspects of mission support, such as, e.g., aerial refueling. For modern militaries engaged in long-range air campaigns or utilizing air forces for power projection, aerial refueling is a core capability providing numerous tactical and strategic benefits,\footnote{\cite{airforce2024airmobility}.} and one where AI might be fruitfully engaged. 

The execution of aerial refueling (encompassing formation flight, approach, coupling, and precise station-keeping) constitutes a highly critical and cognitively demanding operation,\footnote{\cite{mao2008dynamics}.} with success necessitating tight coordination between the tanker aircrew, the refueling operator, and the receiver pilot.\footnote{\cite{nato2013refuel}.} This procedural complexity is further compounded by environmental variables, visual degradation during night operations, and physiological factors such as fatigue during extended duration sorties.\footnote{\cite{jha2005air}.}

To mitigate these challenges, states and industry partners have explored ways to automate (parts of) the refueling process, introducing concepts such as autopilots and visual tracking as early as the 2000s.\footnote{\cite{valasek2005vision}.} More recently, unmanned aerial vehicles have been proposed for taking over the role of tankers, with manned-unmanned refueling\footnote{\cite{boeing2024validates}.} being facilitated through the deployment of systems like the Boeing MQ-25 Stingray\footnote{\cite{boeing2026stingray}. Such efforts are also in line with the European Union's (EU) intent to develop a more resilient air-to-air refueling capability for EU member states (\cite{pesco2026future}).} or fully autonomous refueling seeing operation through systems like the Airbus Auto'Mate.\footnote{\cite{airbus2023achieves}.} Using AI for navigation and cooperative control,\footnote{\cite{airbus2023aerial}.} Airbus has demonstrated a capability for autonomous refueling of multiple autonomous aircraft in a single operation, and Boeing seems intent on replicating similar capabilities for the United States defense establishment.\footnote{\cite{boeing2025autonomy}.}

AI-enabled and autonomous systems developed for mission support roles such as these will generally not raise many ethical or legal challenges, with the majority of concerns being of a strongly technical nature; essentially, ``Can the system do what is required of it?''. However, if algorithmic systems are developing flight plans and coordinating groups of aircraft operating in close proximity, there will be concerns about how well these systems take into account potential dangers to civilian aircraft or civilians on the ground. This might be addressed by simple operational limits or deployment parameters. However, it is critical to be clear that this is a real potential concern, and one which would need to be addressed through either technical or deployment solutions. 

\subsection{Research and Development}\label{subsec_optimization_r&d}

Numerous companies tout the potential ability of AI to transform and improve academic and engineering research and development.\footnote{E.g., \url{https://elicit.com/}; \url{https://consensus.app/}; \url{https://www.perplexity.ai/encyclopedia/researchers}; \url{https://scite.ai/}; \cite{bellan2025sam}. See also \cite{grossmann2023ai,vannoorden2023ai}. Cf. \cite{chubb2022speeding} for a more balanced assessment.} While there are good grounds to be wary of this claim outside of reference to very narrow AI systems,\footnote{\cite{roose2025where}.} within the military, AI systems have shown promise as a means to conduct wargames and derive actionable information from them. 

%
While once restricted to simple scripted behaviors, wargames and military simulations have evolved to include complex adaptive systems.\footnote{\cite{zaptech2025strategic}.} In the use of AI for wargaming, the United States is the global leader, with the Defense Advanced Research Projects Agency (DARPA) and the Air Force pushing ``Game Theory'' AI. For example, the DARPA ``Gamebreaker'' Program, launched in 2020, focused on using AI to ``break'' simulations;\footnote{\cite{moss2020gamebreaker}.} rather than just playing a game, AIs were tasked with finding imbalances or exploiting rule sets, with the goal being to discover unconventional strategies human commanders might overlook. Using commercial games (like StarCraft II or Command: Modern Operations) as testbeds to train AIs to exploit strategic flaws,\footnote{\cite{csi2022darpa}.} the program has led to follow-up work with industry partners.\footnote{\cite{mckeon2025northrop}.}

Across the Atlantic, the Digitalization and Technology Research Center of the German Bundeswehr has backed Ghostplay, a multi-institutional project\footnote{\cite{ghost2026tactical}.} which aims to create high-performance synthetic environments to simulate tactical decision-making, specifically for air defense and the suppression of enemy air defenses (SEAD). Using AI to control swarms of assets,\footnote{\cite{dtec2024ghost}.} the enemy AI might simulate a drone swarm attack, while the friendly AI attempts to optimize air defense responses in real-time, focusing on ``fight-at-machine-speed'' capabilities.

Focusing on interoperability between partners, NATO has also ramped up its wargaming via the WIN conference series, using tools like WISDOM (Wargaming Interactive Scenario Digital Overlay Model).\footnote{\cite{nato2025concept}.} NATO is also experimenting with generative AI (GenAI) to assist in red teaming -- using AI to generate realistic diplomatic or hybrid warfare injections during a game to test how alliance members react to political ambiguity.\footnote{\cite{nato2025harnessing}.}

The shift from scripted behavior to adaptive AI in projects like GhostPlay and Gamebreaker relies on Deep Reinforcement Learning (DRL)\footnote{\cite{dtec2024ghost}.} and likely also Multi-Agent Reinforcement Learning (MARL).\footnote{\cite{borchert2022free}, esp. p. 17.} These emerging techniques bring promises of better analysis and decision-making, but they also raise potential risks as well. DRL systems do not just follow rules, they ``learn'' strategies and approaches by playing millions of simulated games and optimizing moves across those simulations. This allows such systems to discover strategies a human may never consider, if only because a human does not have the time to play millions of iterations of a game in order to discover such novel approaches. However, that very ability to discover and develop novel moves, coupled with their stochastic nature, makes these systems potentially unexplainable and unpredictable, which, while being part of their allure, also represents a lurking danger. 

MARL systems can suffer from the same type of benefit and risk, namely of being potentially unpredictable due to how they are trained. This might also be compounded by multiple systems acting in concert as a swarm, making behaviors not just unpredictable to humans due to the training regimen, but also because swarm behavior may be alien to human reasoning. By focusing on cooperative distributed decision-making, these systems may be optimized for modern drone warfare but also fully and fundamentally opaque to humans, raising basic ethical, legal, and military challenges to utilizing these systems responsibly.\footnote{See Section \ref{subsec_autonomousengagement_swarms} above, as well as e.g., \cite{miller2019explanation,robbins2019misdirected,robbins2025when,langer2021what,maclure2021ai,vredenburgh2021right,peters2022explainable} for various challenges related to opacity. For a response to some such challenges, see, e.g., \cite{ross2024ai}.}

Finally, utilization of generative AI, which is apt to imply at least some use of LLMs, potentially raises numerous and significant concerns relating to privacy, security of operations, reliability of outputs, and repeatability of any tests or strategies. Any use of GenAI, and especially LLMs, is thus apt to raise a host of concerns related not just to ethics (e.g., privacy) and law (e.g., illegal use of data for model training), but also basic security concerns about the risks of deploying these systems at scale in domains where information can be the difference between life and death.

\section{Military Bureaucracy}\label{sec_bureaucraticAI} 

When thinking about militaries, it is easy to focus on the ``sharp'' end of the institution and lose sight of just how much seemingly mundane effort is necessary for the fighting arm of a military to function. But all of the less ``flashy'' work is necessary, and an effective bureaucracy is a core component of an effective military (or any other institution). Militaries must also navigate complex interconnected systems of law, policy, and doctrine, discerning what they may and must do by examining volumes of code and regulation. For both the navigation of policy and for streamlining the everyday bureaucratic elements of military planning and decision-making, AI tools have been developed and marketed to improve efficiency and outcomes across military institutions. However, all of these raise potential challenges as well. 

\subsection{Policy Guidelines, Human Resources, and Acquisitions}\label{subsec_bureaucraticAI_HR}

At a minimum, virtually every state's military will be subject to international law, its own national laws, and potentially local laws where bases are operated, as well as policy and doctrine from the state's central government and from the precise military institution making decisions at any given moment (e.g., the army, navy, or air force), and all of these rules impact on operations, human resources management, acquisitions, and more. 
In response to such a dizzying array of rules, the United States Department of Defense (US DoD)\footnote{Until and unless the name is changed by statute by the United States Congress, the authors will not use a different name for the United States Department of Defense.} developed the GAMECHANGER application, a tool for searching and filtering through the various documents guiding US DoD operations. GAMECHANGER also includes tools to identify policy gaps, duplications, and conflicts,\footnote{\cite{baily2025gamechanger}; \myurlgamechanger.} critical tasks given that ``if you read all the Department of Defense's policies, it would be the equivalent of reading through `War and Peace' more than 100 times''.\footnote{\cite{dia2022gamechanger}.} Some have maligned that the U.S. military needs an AI to ``make sense of its own `byzantine' and `tedious' bureaucracy'',\footnote{\cite{klippenstein2023pentagon}.} but, even granting there are too many rules and regulations, the simple fact is that militaries by their nature must navigate numerous and potentially competing bases of governance. 

GAMECHANGER provides a solution to this challenge, allowing personnel not just to utilize an AI tool, ``a large language model-based application'',\footnote{\cite{baily2025gamechanger}.} to search for relevant documents, but also to see what other documents might impact on their current tasks. This can improve efficiency greatly by simply shortening the amount of time needed for searching for policy, as well as by quickly presenting personnel with competing policy which may require their judgment in resolving conflicts. These bureaucratic improvements can lead to improvements in managing human and other resources, in streamlining acquisitions, and in ensuring that operations are done in accordance with all relevant ethical, legal, and doctrinal guidelines. As GAMECHANGER utilizes an LLM-based system primarily for queries and retrievals, and less for production of texts, it also avoids many of the pressing issues relating to LLMs and their inability to track truth. Thus, GAMECHANGER, by providing a very narrow use case for a specific AI system, is able to meet a real need without raising many concerns. 

That being said, use of AI, in particular LLMs, in relation to policy guidelines can introduce risks, especially if done without care. As policy guidelines encapsulate legal and other requirements, it can be incredibly important to have them \textit{exactly} right, but many LLMs and other AI systems may flatten guidelines, omit key terms, or alter key terms which have import. For example, consider how carefully any legal document is written and how much may hang on individual terms, each of which may track to whole swaths of statutory, treaty, and case law. If such terms are changed to semantically -- but not legally -- similar terms by an AI system, this could result in illegal and potentially even dangerous actions. GAMECHANGER appears to avoid these risks by careful limitations to how it is deployed in the US DoD and across partner institutions, and given the complexity of rules governing singular national militaries, and the fact that many national militaries are entrenched members of overlapping alliance structures potentially with their own additional guidelines, such an AI tool makes much sense. However, any development and deployment of such tools must be undertaken with an eye to the risks in doing so without due care. 

\subsection{LLMs, Emails, and the Military as a Corporate Body}\label{subsec_bureaucraticAI_corporatestreamlining}

In addition to navigation of policy, the military, like any corporate body, must carry out a vast number of everyday tasks such as reading and responding to emails, finding information requested by internal and external partners, creating reports for internal use and for dissemination to outside offices, and much more. For such everyday bureaucratic and corporate tasks, numerous companies have offered AI-powered solutions. In the United States, the Ask Sage platform was launched to help US DoD and other government employees in a range of tasks where generative AI may prove useful. Utilizing various AI systems, Ask Sage lists coding, acquisition, compliance, data analysis, and automation as potential use cases for the system. Ask Sage is claimed to be able to provide all these capabilities by leveraging numerous different AI capabilities, in particular, LLMs and agent-based systems that can act independently to carry out tasks. Users need only to, ``through a conversational interface, ask a question or submit a task to be completed'',\footnote{\url{https://www.asksage.ai/platform/}.} and Ask Sage begins to help. In Germany, the iFinder system fills a similar role, taking on some of the capabilities of GAMECHANGER (search and highlighting of relevant documents) and combining them with the GenAI tools offered by platforms like Ask Sage. The German Bundeswehr already uses this tool,\footnote{\cite{intrafind2025ifinder}.} and Intrafind, the company offering iFinder, claims that because the system is ``fed'' with validated data -- a central point of iFinder is that organization-internal data is central to the platform -- the LLM chatbot cannot ``hallucinate'' (fabricate) like ChatGPT and related public offerings are known to do.\footnote{\cite{intrafind2026datenblatt}, translated from German by the authors.}

However, while streamlining the everyday bureaucratic elements of a military's functions surely represents a boon, these systems raise numerous and, in some cases, insurmountable issues. First, all of the general issues of LLMs are implicated for these systems, from deskilling to hallucinations, from faulty analysis to unpredictable outputs.\footnote{For discussion of some of these problems and others, see, e.g., \cite{hadi2023large,biddle2024meta,cappelli2024will,hicks2024chatgpt,wierda2024when,coveney2025wall,guest2025against,karbasi2025impossibility,kosmyna2025brain,kwik2025digital,maiberg2025microsoft,suzuki2025people,voinea2025sorrows,robbins2026losing}} Intrafind's statement that iFinder presents no risk of hallucinations is also almost certainly false, as so-called hallucinations are not a bug of LLMs, but a central feature; these systems do not have models of the world and are not tethered to truth or falsity, instead only generating most plausible next tokens in response to a query.\footnote{\cite{hicks2024chatgpt}.} Thus, \textit{all outputs are hallucinations}, or fabrications, and it just happens to be that many fabrications are also true because probable next tokens will often be aligned with what is the case.\footnote{\textit{Ibid}.} If iFinder is entirely trained and operated using only verified and validated data, this would go some way to improving the security of the system and potentially also the veracity of its outputs, but outputs will still be approximations of most likely next tokens in response to a request, and certainly not anchored to truth and possibly not even to the contents of organization documents. 

LLMs, due to the ways they are trained -- and likely must be trained -- also present potentially significant security risks.\footnote{For varying risks, see, e.g, \cite{biddle2024meta,kwik2025digital,reddy2025echo,schulz2025echgram,souly2025poisoning}.} Given the volume of data needed to train these systems, virtually all rely, to some extent, on scraping the public internet for data. But the internet may be populated with vast quantities of poisoned data, where adversarial actors have inserted data into text, images, audio, or other files which may compromise the effectiveness of a system or potentially even create backdoors into a system when certain triggers are met. Moreover, recent research shows that as few as 250 instances of poisoned data can be enough to compromise a system,\footnote{\cite{souly2025poisoning}.} making virtually all LLMs likely at risk. Using LLMs, even for seemingly mundane back-office tasks, is thus highly risky, as everyday bureaucracy is massively impactful for military planning and decision-making. 

\section{Improving the Debates Going Forward: Specificity}\label{sec_specificity} 

AI is used in natural language processing, object identification, pattern recognition, route planning, search engine optimization, and many other applications. This is because, as noted at the outset of this article, ``AI is not a discrete technology, like railroads or airplanes. It is a general-purpose enabling technology, like electricity, computers, or the internal combustion engine, with many applications''.\footnote{\cite{scharre2023four}, p. 3.} Yet despite this, many ongoing debates center around ``AI'' as though this were a single technology, with ``military AI'' being similarly invoked without further specification.\footnote{Even in research focusing on the importance of communicating the realities of AI-enabled systems -- e.g., \cite{davidovic2025rethinking} -- one can find imprecise language of ``AI tools'' or ``AI weapons''.}

The above survey of AI technologies in use across the military demonstrates the folly of such a unitary view. Our survey should not be taken to be exhaustive or to provide a fully-fledged demarcation of different types of systems and the taxonomic lines that may separate them.\footnote{This is an important task which has a potential for significant impact on how we approach the engineering, regulation, and ethical and effective use of such technologies. However, such taxonomic work will necessitate involved scientific and philosophical treatment which is outside the scope of the arguments being presented here.} The attendant discussion of risks, challenges, and issues related to these various AI(-enabled) systems should likewise not be treated as necessarily complete. Further, our intent has not been to show that any specific AI or AI-enabled system ought (not) be used within the military or any other domain. Rather, we hope to have shown just how many uses of AI there are, how varied AI and AI-enabled systems may be, and how distinct the ethical, legal, and operational challenges may be due to the exact nature of some AI(-enabled) system and its intended use. The central finding of this article is thus a higher level point relevant for the ongoing debates around ``military AI'' -- and indeed AI in numerous domains -- namely that the debates ought, as much as possible, to eschew the increasingly unhelpful and radically imprecise talk of ``AI'' as such. 

While certain aspects of AI development and training may be germane to numerous overlapping AI approaches, these also will not necessarily hold across the board, as ML systems will not be developed in the same way, offer the same opportunities, or present the same challenges as, say, classical symbolic AI (sometimes referred to as ``good old-fashioned AI'', or GOFAI). Furthermore, many ``artificial intelligence'' systems do not possess ``intelligence'' of any sort -- at least, intelligence as traditionally understood -- yet the term can color how one views the capabilities of a system, creating risks. For these reasons, we should speak of large language models, natural language processing systems, computer vision systems, big data analytics, and the like, rather than leaning on the familiar but less useful ``AI''. 

More practically speaking, the potential problems and challenges of a decision-support system used to process and analyze data are entirely distinct from those related to a large language model deployed to streamline back-office bureaucracy.\footnote{Note though that a non-LLM system might include an LLM within it for distinct functions, complicating the operational, ethical, and legal picture.} Likewise, the ethical, legal, and operational worries attending AI systems used in mission support and research and development will be different still. More starkly, an entire range of AI processes may be embedded into autonomous weapon systems, raising a veritable host of distinct concerns relating partly to AI and partly to the autonomous nature of AI-AWS. Going forward, it is thus critical that discussions be clear about whether they concern, e.g., AI analysis tools for AI-DSS, AI-enabled object identification used for autonomous targeting, or AI in the form of LLMs used to provide advisement or bureaucratic streamlining or simplification/transmission of orders -- each of which itself raises distinct problems. Artificial intelligence is not \textit{a technology} but rather a \textit{class of technologies}\footnote{\cite{corea2019ai,metcalfe2021systemic,google2025ai,shaw2025revised}.} and treating the problems of this whole class as co-extensive with each particular system is almost certain to generate confusion, failures in communication, and, in the worst case, regulations and governance structures which are unworkable or downright foolish. 

Moreover, while we have focused on AI within the military as a case study, it should be stressed that the same holds true for AI deployment across virtually every other domain. And it is instructive to see just how broad ``AI'' is within the military alone, as it highlights that issues must be tackled in a piecemeal and specific fashion, and not with an eye to ``AI issues'' as some broad call to arms. Each system brings its own challenges, just as each domain has its own challenging sub-areas, and by seeking to lump together all systems under the (very wide) umbrella of ``AI'', we are apt to miss domain- and system-specific challenges \textit{and opportunities}. At a time when everything is hailed as being or having ``AI'', it is critical that we drop the act and be precise about what exactly we have in mind, that we set definitional and argumentative baselines that ``will support intellectual discipline''.\footnote{\cite{tucker2022artifice}.} Because the challenges, ethical and legal hazards, operational risks, and potential solutions will not be found by looking to some amorphous notion of artificial intelligence. Instead, they will be found by examining the nuts and bolts of real systems in real environments, designed and used by real people. 

\pagebreak 


\section*{List of abbreviations}

\begin{tabular}{l|l} 
    \bf Abbreviation & \bf Meaning \\ \hline \hline
    AI & Artificial Intelligence \\ \hline
    AI-DSS & AI-Enabled Decision Support Systems \\ \hline
    AI-AWS & AI-Enabled Autonomous Weapon Systems \\ \hline
    ATGM & Anti-Tank Guided Missile \\ \hline
    AWS & Autonomous Weapon Systems \\ \hline
    CCA & Collaborative Combat Aircraft \\ \hline
    DARPA & Defense Advanced Research Projects Agency \\ \hline
    DRL & Deep Reinforcement Learning \\ \hline
    DSS & Decision Support Systems \\ \hline
    EU & European Union \\ \hline
    EW & Electronic warfare \\ \hline
    GenAI & Generative Artificial Intelligence \\ \hline
    GOFAI & Good Old-Fashioned AI (also known as ``symbolic AI'') \\ \hline
    IDF & Israel Defense Forces \\ \hline
    ISR systems & Intelligence, Surveillance, and Reconnaissance systems \\ \hline
    LLMs & Large Language Models \\ \hline
    LRASM & Long-Range Anti-Ship Missile \\ \hline
    MARL & Multi-Agent Reinforcement Learning \\ \hline
    ML & Machine-Learning \\ \hline
    NATO & North Atlantic Treaty Organization \\ \hline
    NLP & Natural-Language Processing \\ \hline
    OODA loop & Observe-Orient-Decide-Act loop \\ \hline
    PANDA & Predictive Analytics and Decision Assistant \\ \hline
    SEAD & Suppression of Enemy Air Defenses \\ \hline
    US DoD & United States Department of Defense \\ \hline
    WISDOM & Wargaming Interactive Scenario Digital Overlay Model \\
\end{tabular}

\pagebreak 

{\small \bibliography{master}}

\end{document}